\documentclass{nature}
\usepackage{times}
\usepackage{xcolor}
\usepackage{graphicx}
\usepackage{amsmath}
\usepackage{amssymb}
\usepackage{mathtools}
\usepackage{amsthm}
\usepackage{physics}
\usepackage{gensymb}
\usepackage{color}
\usepackage{url}
\usepackage{tabularx}
\usepackage{soul}
\usepackage[final]{changes}
\makeatletter
\@namedef{Changes@AuthorColor}{red}
\colorlet{Changes@Color}{red}
\makeatother
\definechangesauthor[name={Added}, color=red]{new}
\definechangesauthor[name={Deleted}, color=blue]{old}

\title{Physics-aware Complex-valued Adversarial Machine Learning in Reconfigurable Diffractive All-optical Neural Network}

\author{Ruiyang Chen,$^{1,\dagger}$ Yingjie Li,$^{1,\dagger}$ Minhan Lou,$^{1}$ Jichao Fan,$^{1}$ Yingheng Tang,$^{1,2}$ Berardi Sensale-Rodriguez,$^{1}$ Cunxi Yu,$^{1,\ast}$ and Weilu Gao$^{1,\ast}$}

\begin{document}

\maketitle

\begin{affiliations}
 \item Department of Electrical and Computer Engineering, University of Utah, Salt Lake City, UT 84112, USA
 \item Department of Electrical and Computer Engineering, Purdue University, West Lafayette, IN 47907, USA
\end{affiliations}
\noindent $^\dagger$These authors contributed equally.\\
\noindent $^\ast$To whom correspondence should be addressed; E-mail: cunxi.yu@utah.edu; weilu.gao@utah.edu.

\pagebreak
\begin{abstract}

Diffractive optical neural networks have shown promising advantages over electronic circuits for accelerating modern machine learning (ML) algorithms. However, it is challenging to achieve fully programmable all-optical implementation and rapid hardware deployment. Furthermore, understanding the threat of adversarial ML in such system becomes crucial for real-world applications, which remains unexplored. Here, we demonstrate a large-scale, cost-effective, complex-valued, and reconfigurable diffractive all-optical neural networks system in the visible range based on cascaded transmissive twisted nematic liquid crystal spatial light modulators. With the assist of categorical reparameterization, we create a physics-aware training framework for the fast and accurate deployment of computer-trained models onto optical hardware. Furthermore, we theoretically analyze and experimentally demonstrate physics-aware adversarial attacks onto the system, which are generated from a complex-valued gradient-based algorithm. The detailed adversarial robustness comparison with conventional multiple layer perceptrons and convolutional neural networks features a distinct statistical adversarial property in diffractive optical neural networks. Our full stack of software and hardware provides new opportunities of employing diffractive optics in a variety of ML tasks and enabling the research on optical adversarial ML.

\end{abstract}

\pagebreak

\section*{Introduction}
\vspace{-20pt}

High-throughput and energy-efficient processing of machine learning (ML) algorithms enables applications in diverse disciplines, including computer vision and autonomous driving\cite{LeCunEtAl2015N,GoodfellowEtAl2016,RodriguesEtAl2021NP}, the discovery of materials and molecules\cite{ButlerEtAl2018N,SeniorEtAl2020N}, and chip and circuit design\cite{MirhoseiniEtAl2021N}. However, with the end of Dennard scaling and Moore's law, the power consumption and integration density of electronic circuits have started to hit a bottleneck of processing algorithms with trillions of arithmetic operations. On the other hand, there is a noticeable increase in security threats on a variety of ML systems\cite{HuangEtAl2011,KurakinEtAl2016APA,FinlaysonEtAl2019S,SunEtAl2020}, which has lead to a widespread effort of developing techniques to detect and respond to adversarial ML threats in real-world applications\cite{atlas}.

Recently, photonic neuromorphic processors are emerging as high-performance hardware ML accelerators by leveraging fundamentally different particles, photons, to break electronic bottleneck thanks to the extreme parallelism from the weak interaction and multiplexing of photons as well as low static power consumption\cite{ShenEtAl2017NP,LinEtAl2018S,HamerlyEtAl2019PR,ZhouEtAl2021NP,GaoEtAl2021APR,LiEtAl2021SR,WangEtAl2022NC}. Particularly, three-dimensional (3D) free-space diffractive optical neural networks, which exploit the out-of-plane dimension for light routing and can potentially host millions of compact active devices and computing neurons in deep network architectures, physically perform the multiplication and addition operations central to ML algorithms through spatial light modulation and diffraction, and demonstrate their capability of performing image classification\cite{LinEtAl2018S,ZhouEtAl2021NP}. Since the demonstration of utilizing 3D-printed passive diffractive components in the terahertz range\cite{LinEtAl2018S}, the advanced compact meta-photonic components in a more readily accessible wavelength ranges, including visible and near-infrared light, are emerging to improve the hardware efficiency and functionality\cite{LeonardEtAl2021P,LuoEtAl2021APA,ZhengEtAl2022APA}. However, it is challenging to adjust diffractive layers once they are fabricated, which makes adaptive training and active learning on physical systems impractically time-consuming. 


Active spatial light modulators (SLMs), on the other hand, become a promising candidate for realizing the reconfigurability in diffractive optical neural networks for better versatility, efficiency, and robustness\cite{ZhouEtAl2021NP}. However, the state-of-the-art demonstration utilizes a phase-only SLM with a $2\pi$ tunable range for diffractive layers, and optical-to-electrical and electric-to-optical conversions are involved between layers. The inevitable loss of phase information during these conversions leads to a substantial measurement deviation from the computer-developed model. Additional adjustment and iterative training for error corrections are needed\cite{ZhouEtAl2021NP,WrightEtAl2022N}, and the reconfigurability and transfer to other tasks are still laborious to accomplish. Multiple conversions between electrical and optical domains also weaken the advantages of employing optical approaches over electrical ones. Moreover, due to the lack of fully reconfigurable and deployable optical hardware and corresponding accurate physical software framework, the understanding of the adversarial ML properties in such systems from both algorithmic and experimental perspectives is challenging and elusive. 

Here, we demonstrate complex-valued, reconfigurable, diffractive all-optical neural networks (CRDONNs) in the visible range based on cascaded multiple layers of cost-effective transmissive SLMs made from twisted nematic liquid crystals (TNLCs), which feature the coupled modulation of transmitted light amplitude and phase. We create a high-throughput, physics-aware training framework, enabling the direct, accurate, and fast deployment of software-generated models onto experimental hardware. With the developed full stack of software and hardware, we theoretically analyze and experimentally demonstrate adversarial attacks on CRDONNs, which are generated from a physics-aware complex-valued fast gradient sign method (C-FGSM). Compared to conventional multiple layer perceptrons (MLPs) and convolutional neural networks (CNNs), the CRDONNs display distinct statistic adversarial ML properties. The demonstrated all-optical ML system not only offers a new route toward the manufacturing and rapid deployment of large-scale photonic neuromorphic processors by leveraging the mature liquid crystal technology and the developed physics-aware calculation framework, but also builds a playground of exploring new adversarial and defense algorithms in an uncharted area of complex-valued neural networks. 

\section*{Results}
\vspace{-20pt}

\noindent\emph{\underline{Complex-valued reconfigurable diffractive all-optical neural networks (CRDONNs)}}

\noindent Figure\,\ref{fig:arch}a describes the architecture of CRDONNs. In our system, there are four TNLC-based transmissive SLMs with the first one for programmable input images and the rest three for reconfigurable diffractive arrays (RDAs). As a laser beam is incident onto SLMs, the optical diffraction occurs on each pixel of SLMs and the diffracted light connects one pixel from a layer to a group of pixels in the next layer. Each pixel on each SLM at location $(x,y)$ has a complex-valued electric field transmission coefficient $t(x,y,V)\mathrm{e}^{j\phi(x,y,V)}$, where $t(x,y,V)$ describes the amplitude modulation as a function of external stimulus $V$ such as applied voltages or grey levels, and $\phi(x,y,V)$ describes the corresponding phase response. When the laser beam passes through the transmissive SLMs, the light intensity and phase are modulated in a coupled manner and the complex-valued modulation can be configured by adjusting the input and output light polarization states\cite{MarquezEtAl2001OE}. With a proper combination of driving voltages or grey levels on SLMs, input images, such as handwritten digits in the \texttt{MNIST} dataset, can be converged to one of pre-defined regions on a camera. Each region represents the output label of input images, and the label associated with the detector region with a maximum intensity is selected and classified for the corresponding input image.

Specifically, as shown in Fig.\,\ref{fig:arch}b, a laser diode with a beam at center wavelength $532\,$nm and having a diameter $\sim4\mathrm{mm}$ is incident onto the image SLM (SLM0). A pair of polarizers (P0 and A0) with orthogonal transmission axes are equipped before and after SLM0 to configure the input and output light polarization states for binarized input images. As a result, the SLM0 is in an intensity mode with a strong modulation of light transmittance. At the output of SLM0, a half-wave plate is used to reconfigure the input polarization state for the next SLM (SLM1). By selecting the output polarization state, the SLM1 is in the phase mode and a complex-valued modulation with a large phase tuning range can be achieved. This combination of a half-wave plate, a SLM, and a polarizer form a RDA, which can be cascaded into multiple layers for diffractive deep neural networks. In our demonstration, we have three RDA layers and at the end there is a camera for capturing the diffraction image after the laser passes through the system.

The adversarial images are generated through a gradient-based process (dashed line in Fig.\,\ref{fig:arch}b)\cite{GoodfellowEtAl2014APA}. Generally, the input image ($I_p$) is first predicted through the CRDONNs system and the loss of the prediction ($\mathcal{L}(I_p,I_t)$) is computed based on the true class label ($I_t$). The sign of the gradient is then computed and the signed gradient is used to construct the output adversarial images. However, the unique complex-valued coupled modulation of light amplitude and phase in the CRDONNs highlights the necessity of tailoring adversarial attack algorithms to match the physical system.

\noindent\emph{\ul{Physics-aware training framework and experimental deployment of CRDONNs}}

\noindent Figure\,\ref{fig:exp_verify}a displays a photo of the CRDONNs experimental setup. The free-space light propagation denoted by green arrows is all-optical. Both the image SLM and RDA-SLMs are electrically reconfigurable and output images are captured by a monochromatic CMOS camera. The detailed description on the experimental setup can be found in \emph{Methods} and \emph{Supplementary Figure\,1a}. Considering the laser beam spot size and the pixel size ($36\,\mu$m$\times36\,\mu$m) of SLMs, we choose a region consisting of $100\times 100$ pixels on the first SLM for input images. We characterized the amplitude and phase modulation responses ($t(V)$ and $\phi(V)$) of both the intensity-mode SLM for input images and the phase-mode SLMs for RDAs as a function of SLM grey level ($V$). We assume each pixel on one SLM has the same response so that $t$ and $\phi$ are not the function of the location $(x,y)$. The amplitude modulation of all SLMs were obtained by measuring the transmitted light intensity using a photodiode. The phase modulation of phase-mode SLMs were measured in an interference experiment and the phase change was calculated based on fringe shift. The phase difference between two grey levels ($0$ and $255$) used to encode input images in the intensity-mode SLM was estimated by matching measured and simulated diffraction patterns. The measured amplitude and phase modulation curves for all SLMs are shown in Fig.\,\ref{fig:exp_verify}b and the phase difference in SLM0 is estimated as $\pi$; see detailed description on the optoelectronic characterizations of SLMs in \emph{Methods} and \emph{Supplementary Figures\,1b--1d}.

The fast and accurate deployment of computer-trained models onto the physical CRDONNs system depends on constructing a trainable and physics-aware framework involving the complex-valued modulation of TNLC-SLMs and the diffraction propagation between layers. In order to incorporate arbitrary device response in RDAs, we employ a device-specific physics-aware training via differentiable discrete complex mapping based on the categorical reparameterization with Gumbel-Softmax (Fig.\,\ref{fig:exp_verify}c)\cite{JangEtAl2016APA,LiEtAl2021}. Conventionally, the $256$ discrete grey levels in experimentally measured device complex-valued modulation response break the gradient chain during the backpropagation in training process. In contrast, the Gumbel-Softmax can approximate the discrete values with the differentiable Gumbel distribution $\sim g(0, 1)$ and the class probability $\theta$. Starting from the loss function of the ML task, the gradient backpropagates through the approximation by Gumbel-Softmax instead of the discrete voltage values directly, and then $\theta$ is updated according to the optimization method. In the forward path of the next epoch, the approximation is first converted to a one-hot vector representing a specific discrete grey level depending on $\theta$, and then converted to corresponding amplitude transmission coefficients and phases by multiplying with experimentally measured modulation curves in Fig.\,\ref{fig:exp_verify}b. As a result, the grey levels on each RDA-SLM become trainable parameters in the Gumble-Softmax algorithm; see \emph{Methods} for more mathematical details. The image SLM encodes input images in a binary manner, where bright and dark pixels corresponds to the complex-valued field transmission of grey levels $0$ and $255$, respectively; see \emph{Methods} for more details on input image encoding. 

In addition, we compare different diffraction models and calculation methods under the framework of scalar diffraction theory\cite{GoodmanEtAl2005PEC} to accurately describe the free-space diffraction in TNLC-SLMs. To leverage the general-purpose graphic computing units (GPUs) for high-throughput calculation and training, the first methodology of calculating diffraction integral is a spectral algorithm based on Fourier transform; see \emph{Supplementary Figure\,2} for an illustration and \emph{Methods} for details. Specifically, we utilize the free-space diffraction impulse function under Fresnel approximation and follows the same pixel grid in SLMs. When the CRDONNs system is trained under Fresnel approximation to create a clear focused spot on the camera for an input handwritten digit $1$ in the \texttt{MNIST} dataset (\emph{Supplementary Figures\,3a and 3b}), the experimental measurement yields a completely random diffraction pattern (\emph{Supplementary Figure\,3d}). This discrepancy can originate from that the calculation grid boundary is not far enough from the experimental boundary of the active region in SLMs, leading to the alias from circular convolution. It is also related to that our pixel size $36\,\mu$m is much larger than the wavelength $532\,$nm, so that the assumption of a uniform phase response from the secondary sources on each SLM pixel becomes inaccurate. 

In contrast, we evaluate the Fresnel-Kirchhoff diffraction by directly calculating integral in a convolutional manner\cite{VdovinEtAl1997}, where the phase integral is calculated based on Fresnel integral and the convolution can also be accelerated through the Fourier transform using GPUs (see \emph{Methods} for detailed mathematical descriptions). As a result, the issues associated with finite boundaries and phase approximations are remedied so that the trainable simulation framework can match experimental results accurately. As shown in \emph{Supplementary Figure\,3c}, the diffraction pattern calculated by incorporating the convolutional Fresnel method in the trained model using the spectral algorithm under Fresnel approximation is completely blurred without any clear focused pattern. However, when the SLM grey levels are trained using the convolutional Fresnel diffraction method, the calculated diffraction patterns from the spectral algorithm and convolutional Fresnel method match experimentally obtained diffraction pattern well (\emph{Supplementary Figures\,3e -- 3g}). 

In addition to the accurate modeling of free-space diffraction, multiple TNLC-SLMs need to be precisely aligned with respect to each other within a range of a few pixels ($100$s'\,$\mu$m); see \emph{Supplementary Figure\,4}. Moreover, although a $100\times100$ region is used in the first image SLM, additional paddings are needed in RDA-SLMs to accommodate the diffraction from edge pixels for accurate match between calculations and experiments in both training and inference processes (\emph{Supplementary Figure\,5a}). When the number of paddings is small, both the simulation diffraction patterns and corresponding intensity percentage in each label detector region with respect to the total intensity in three detector regions show significant deviation from experimental measurements (\emph{Supplementary Figures\,5b and 5c}). The increasing number of paddings display a good agreement between simulation and experiment (\emph{Supplementary Figure\,5d}). We choose the padding size to be $80$ on each side of RDA-SLMs, so that the active region of RDA-SLMs has a size of $260\times260$. 

With the accurate diffraction calculation and precise hardware alignment, the trained grey levels of SLMs can be fast deployed on the CRDONNs experiment setup. Figure\,\ref{fig:exp_verify}d displays a representative input image of a handwritten digit $1$ from the \texttt{MNIST} dataset, experimentally measured diffraction pattern and corresponding intensity percentage in each label detector region, and calculated diffraction pattern and corresponding intensity percentage in each label detector region using the developed physical simulation engine. Both simulation and experimental results agree with each other. Without specific mention, black (white) color indicates large (small) normalized light intensity in all gray images. The blue dashed lines in experimentally measured images correspond to a size of $100\times100$ pixels on the first SLM and the red solid lines correspond to the camera-captured region. Furthermore, we trained the CRDONNs to classify three digits $\{0,1,7\}$ in the \texttt{MNIST} dataset, and the confusion matrices and accuracies obtained from calculations and experiments in Figs.\,\ref{fig:exp_verify}d and \ref{fig:exp_verify}f show excellent agreement ($95\,\%$ accuracies from both calculation and experiments). 

\noindent\emph{\underline{Adversarial attacks on CRDONNs}}

\begin{figure}
    \centering
    \includegraphics[width=0.9\textwidth]{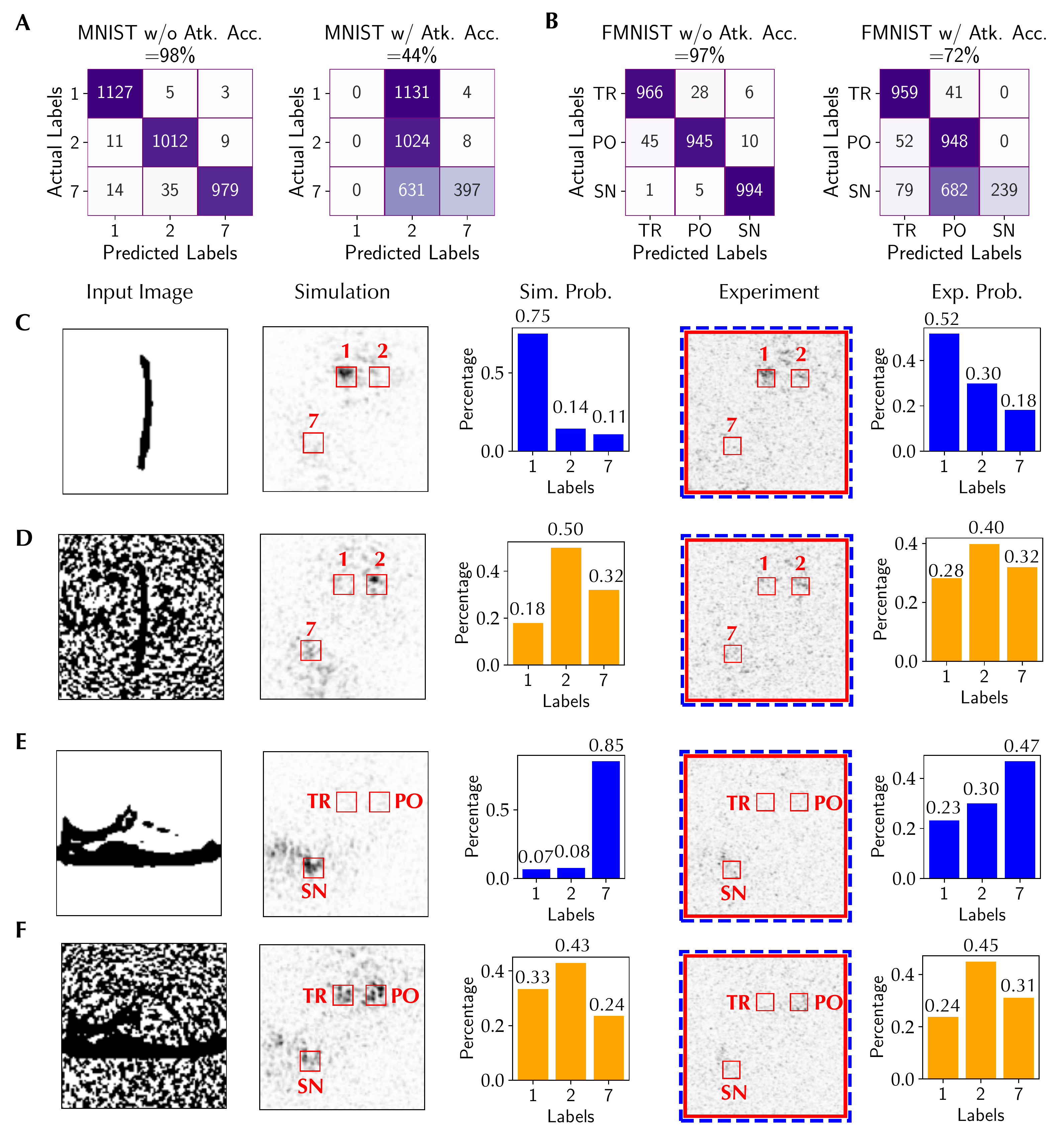}
    \caption{\label{fig:exp_adv} \textbf{Experimental demonstration of adversarial attacks in CRDONNs.} Confusion matrices for (a)\,\texttt{MNIST} and (b)\,\texttt{F-MNIST} datasets before and after attacks generated through a physics-aware complex-valued fast gradient sign method. The size of input images is $100\times100$. (c)\,Input image of a digit $1$ from the \texttt{MNIST} dataset without adversarial attack, (d)\,input image of the same digit $1$ with adversarial attack, (e)\,Input image of a sneaker from the \texttt{F-MNIST} dataset without adversarial attack, (f)\,input image of the same sneaker with adversarial attack, as well as corresponding simulated inference diffraction pattern, the intensity percentage obtained from simulation, experimental diffraction pattern, and the intensity percentage obtained from experimental measurements.}
\end{figure}

\noindent With the developed full stack of physics-aware simulation software framework and fast deployable CRDONN hardware, we demonstrate adversarial attacks produced through a gradient-based C-FGSM algorithm, where the sign of Gumbel-Softmax gradient on the grey level determines the pixel flips in input images; see \emph{Methods} for detailed descriptions of C-FGSM. Figure\,\ref{fig:exp_adv}a shows the confusion matrices of \texttt{MNIST} classification before and after adversarial attacks. The accuracy drops substantially from $98\,\%$ to $44\,\%$, confirming a successful attack. We also demonstrate the attack to the $\{1,2,7\}$ labels in the \texttt{Fashion-MNIST} (\texttt{F-MNIST}) dataset, which are corresponding to trousers (TR), pullovers (PO), and sneakers (SN). The accuracy also drops substantially from $97\,\%$ to $72\,\%$; see Fig.\,\ref{fig:exp_adv}b for confusion matrices. In the \texttt{MNIST} dataset, the generated adversarial images strongly attack the labels of $1$ and $7$ to make them to be classified as the label $2$. In the \texttt{F-MNIST} dataset, the generated adversarial images only attack the sneaker substantially; see \emph{Supplementary Figure\,6} for additional analysis plots. 

Figures\,\ref{fig:exp_adv}c and \ref{fig:exp_adv}d display representative simulation and experimental results of a digit $1$ in the \texttt{MNIST} dataset with and without adversarial attacks. The clear light intensity redistribution confirms a successful adversarial attack of digit $1$ to be mistakenly classified as $2$, when the input image is generated using the C-FGSM algorithm. In addition, the digit $7$ is successfully attacked to be classified as $2$ as well, while the digit $2$ maintains; see \emph{Supplementary Figures\,7, 8} for additional images and data. These observations are consistent with the change of confusion matrices in Fig.\,\ref{fig:exp_adv}a. Similar attacks can also be observed in other models consisting of three different digits in the \texttt{MNIST} dataset, which are summarized in Table\,\ref{table:mnist_acc}. This suggests a general applicability of our C-FGSM algorithm to the \texttt{MNIST} dataset. We further apply this algorithm to the \texttt{F-MNIST} dataset with the same $\{1,2,7\}$ labels ($\{$TR, PO, SN$\}$). As shown in Fig.\,\ref{fig:exp_adv}b and confirmed in Figs.\,\ref{fig:exp_adv}e and \ref{fig:exp_adv}f, the label SN can be successfully attacked to be classified as the label PO. However, the binarization of input images in the \texttt{F-MNIST} dataset distorts object features, and the overall light intensity is weak with adversarial input images. As a result, in experiments, the strongest intensity percentage in the original label may not be completely flipped to another label; see \emph{Supplementary Figure\,9}. There is still a clear reduction of intensity percentage of the original SN label, demonstrating the effectiveness of adversarial attacks. 

\begin{figure}
    \centering
    \includegraphics[width=\textwidth]{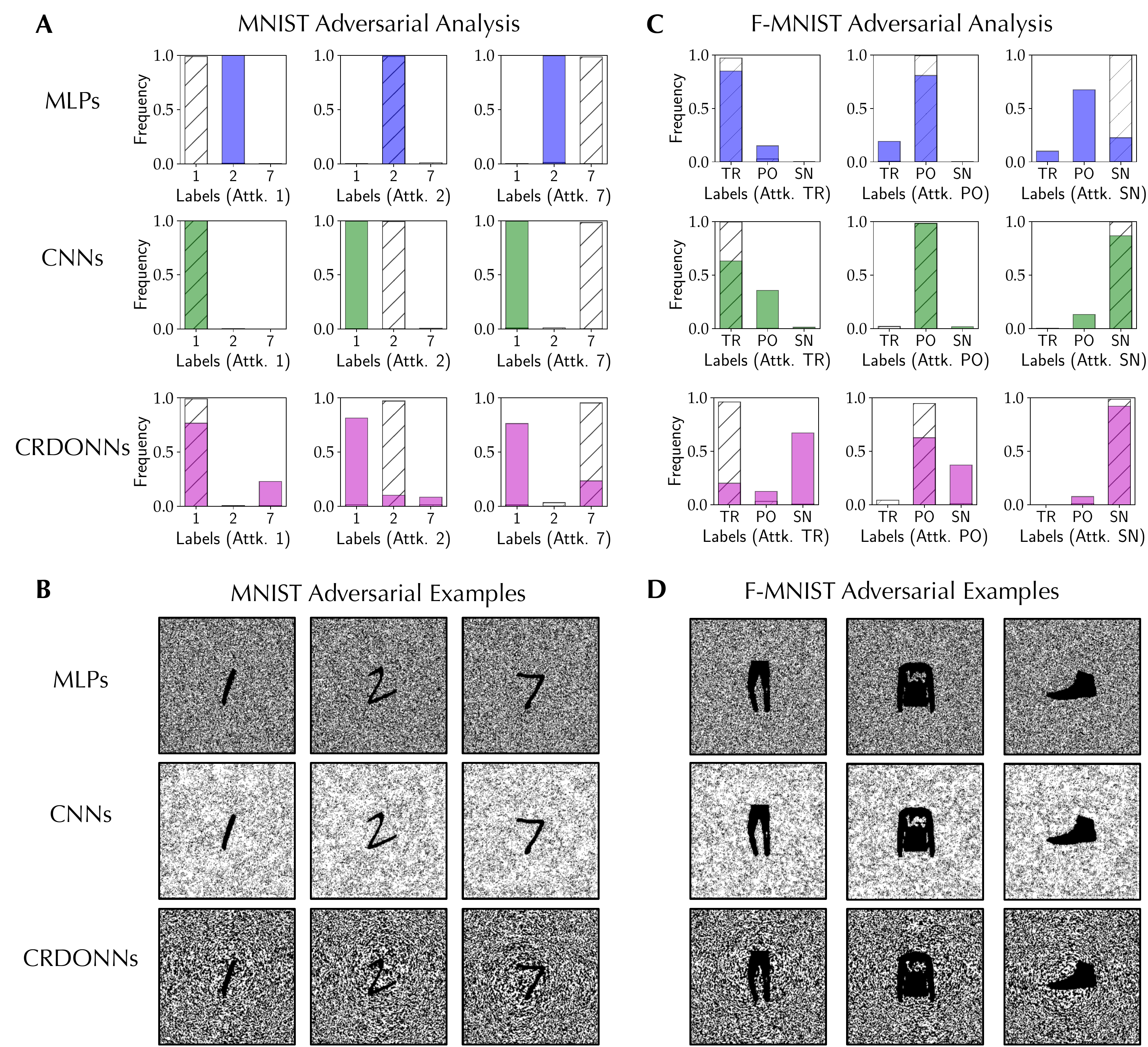}
    \caption{\label{fig:adv_analysis} \textbf{Analysis of adversarial machine learning properties in different neural networks.} (a)\,Classification distribution before and after attacks for the \texttt{MNIST} dataset and (b)\,a few adversarial images. (c)\,Classification distribution before and after attacks for the \texttt{F-MNIST} dataset and (d)\,a few adversarial images. The size of input images is $260\times260$.}
\end{figure}

Finally, we perform a comparative analysis of the adversarial attacks generated from FGSM-like algorithms in various neural networks architectures, including MLPs, CNNs, and CRDONNs; see \emph{Methods} for detailed architectures of MLPs and CNNs. In order to have a fair comparison, the image size for all three architectures is $260\times260$ because of the size of gradients in the RDA-SLMs of CRDONNs. Note that the input image size in CRDONNs experiments is $100\times100$, which is determined by the laser spot size. Figures\,\ref{fig:adv_analysis}a shows the classification distribution of \texttt{MNIST} labels before (bars with forward slash patterns) and after attacks (color bars), as well as a few examples of adversarial input images shown in Fig.\,\ref{fig:adv_analysis}b. In MLPs, the attack on digits $1$ and $7$ is strongly toward the digit $2$, suggesting the robustness of digit $2$ under FGSM-algorithm-generated adversarial attacks. The CNNs architecture displays a strong attack tendency toward and the robustness of digit $1$. In contrast, the behavior of CRDONNs under C-FGSM-generated adversarial attacks in the \texttt{MNIST} dataset is distinctly different from MLPs and CNNs. This highlights the underlying optical diffraction mechanism, which drives the execution of ML tasks in the CRDONNs, could be intrinsically complicated instead of simply mimicking operations in conventional neural networks. Moreover, in the \texttt{F-MNIST} dataset, the behavior of CRDONNs is different from MLPs and CNNs as well; see Fig.\,\ref{fig:adv_analysis}c for the adversarial analysis, \emph{Supplementary Figure\,10} for all corresponding confusion matrices, and Fig.\,\ref{fig:adv_analysis}d for a few adversarial input images. 

\section*{Discussion}

The further improvement of the CRDONNs system can not only handle complicated computer vision tasks with fine features, but also open up windows for better understanding the mathematical representation and capability of general diffractive optical neural networks so that to design advanced attack and defense algorithms in general complex-valued neural networks. For example, the hardware improvement strategy could include the employment of sub-wavelength high-efficiency diffractive units\cite{LiEtAl2019S}, the incorporation of large-size diffractive arrays and deep neural network architectures, and the grey-scale encoding of input images without the image feature loss from binarization. From the software perspective, we can apply adversarial attacks mainly to the phase of input images if they are encoded fully in the complex-valued domain. 

\begin{figure}
    \centering
    \includegraphics[width=0.9\textwidth]{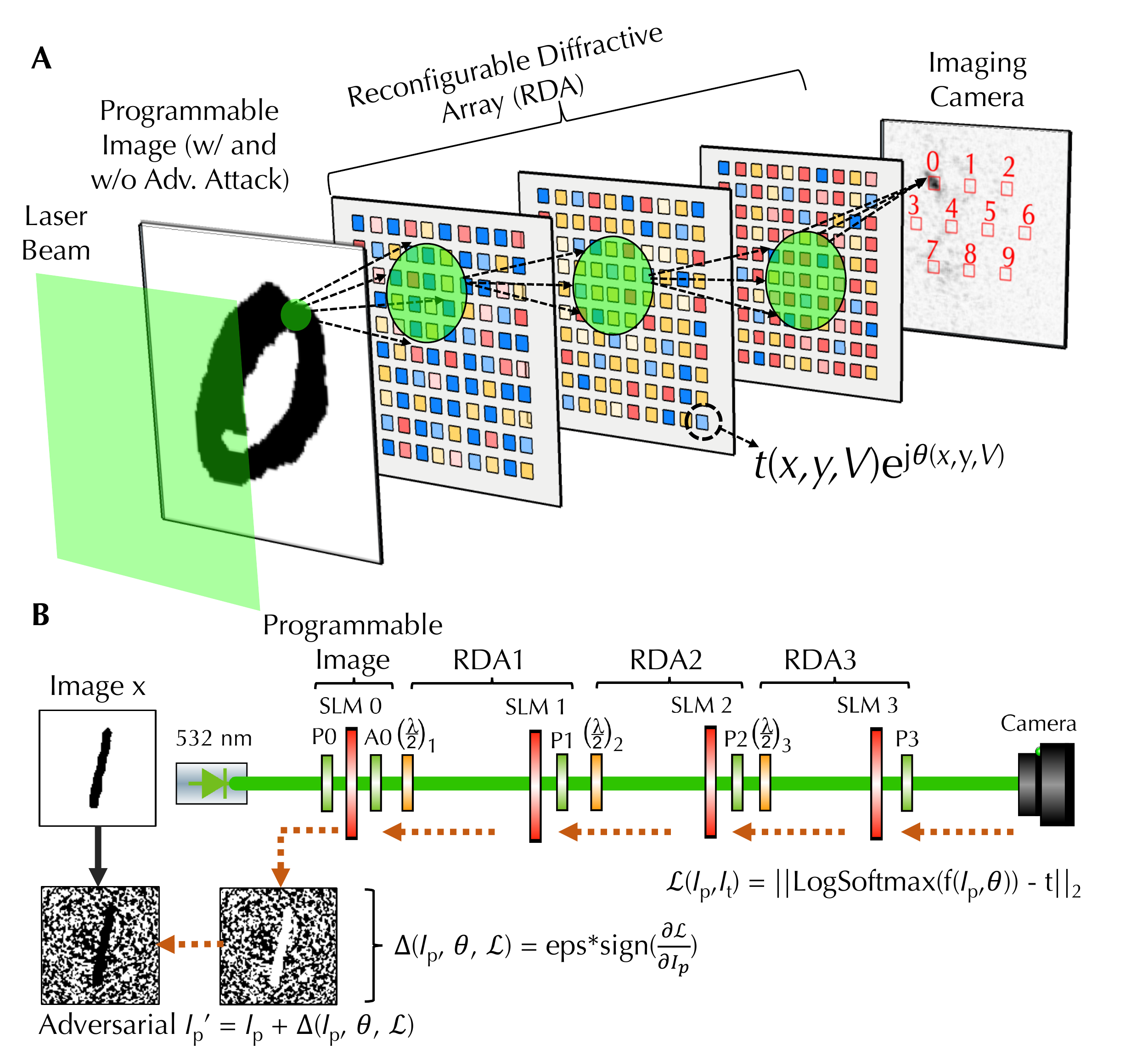}
    \caption{\label{fig:arch} \textbf{Overview of CRDONNs architecture and its adversarial attacks.} (a)\,Schematic diagram of CRDONNs. The reconfigurability of diffractive arrays and input images with and without adversarial attacks are implemented by using the SLMs based on twisted nematic liquid crystals. (b)\,The detailed illustration of CRDONNs and the backpropagated gradient flow to generate adversarial images based on a gradient-based method.}
  \end{figure}

\begin{figure}
    \centering
    \includegraphics[width=0.9\textwidth]{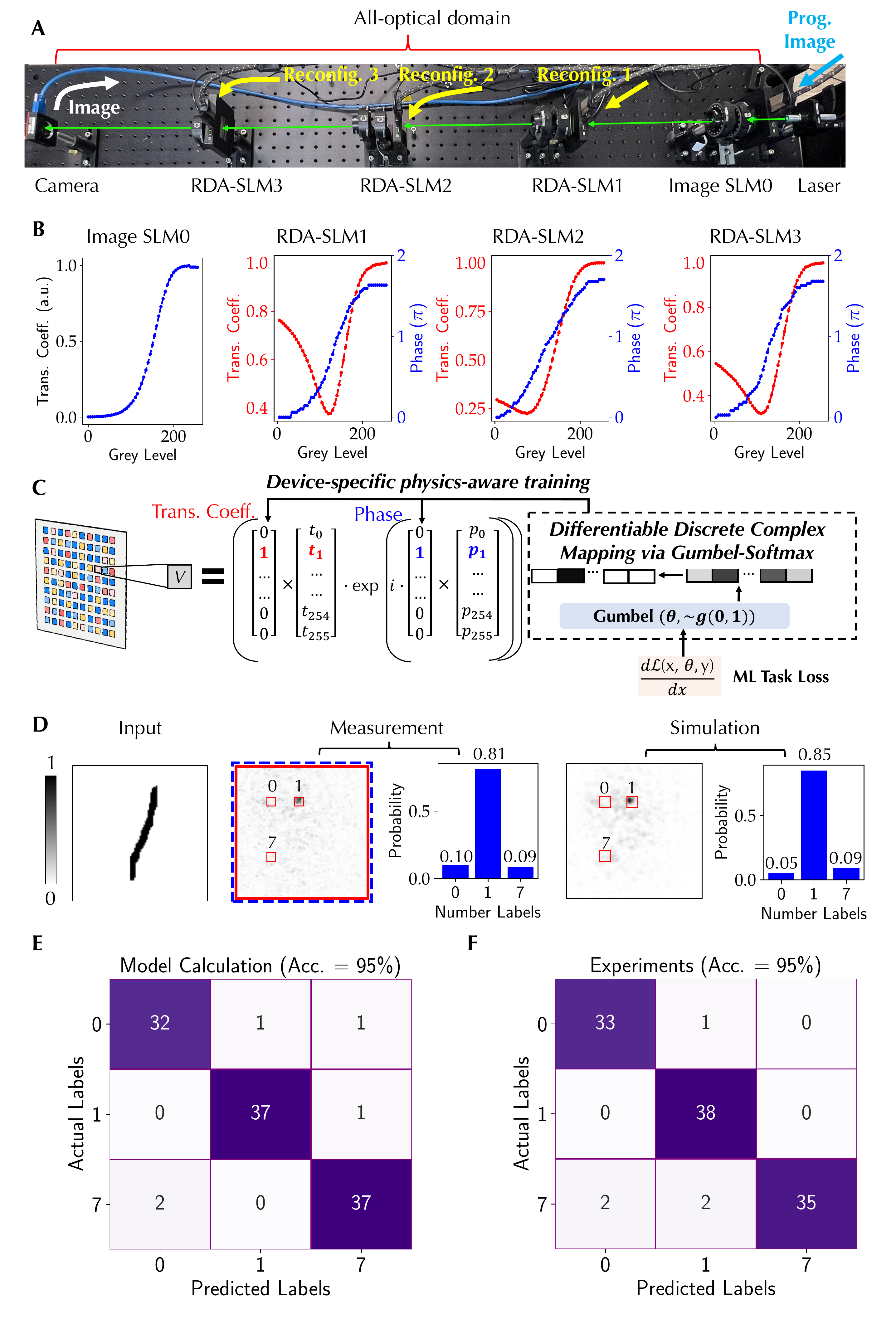}
    \caption{\label{fig:exp_verify} \textbf{Physics-aware software framework and experimental deployment of CRDONNs.} (a)\,A photo of the CRDONNs experimental system. (b)\,The modulation responses of transmitted electric field amplitude and phase for SLMs used for reconfiguring input images and diffractive arrays. (c)\,Illustration of implementing differentiable discrete complex mapping via Gumbel-Softmax for device-specific physics-aware training. (d)\,A representative input image of a handwritten digit $1$ from the \texttt{MNIST} dataset, experimentally measured diffraction pattern and corresponding intensity percentage in each label detector region, and calculated diffraction pattern and corresponding intensity percentage in each label detector region using the developed physical simulation engine. (e)\,Confusion matrix of a computer-trained model consisting of ${1,2,7}$ labels in the \texttt{MNIST} dataset. (f)\,Experimentally measured confusion matrix.}
  \end{figure}

\newpage
\begin{methods}

\subsection{Experimental setup.} The schematic diagram of the experimental setup is illustrated in Fig.\,\ref{fig:arch}b and \emph{Supplementary Figure\,1a}, and the photo is shown in Fig.\,\ref{fig:exp_verify}a. The laser diode (CPS532 from Thorlabs, Inc.) has the center wavelength $532\,$nm. The transmission axes of linear polarizers and the fast optical axes of zero-order half-wave plates are all referenced with respect to the direction normal to the lab bench, and all respective angles are shown in \emph{Supplementary Figure\,1a}. The distance between SLMs and between the last SLM and camera is set as $27.94\,$cm. The polarizers (P0 and A0) before and after SLM0 are configured to have SLM0 to operate with a strong modulation of transmitted electric field amplitude (intensity mode). In contrast, the half-wave plate and polarizer ($(\lambda/2)_1$ and P1) before and after SLM1 are configured to have SLM1 to operate with a strong modulation of transmitted electric field phase (phase mode), together with a moderate modulation of light amplitude. Both half-wave plates and polarizers for SLM2 and SLM3 are configured in the same way. All transmissive SLMs are the LC 2012 model from HOLOEYE Photonics AG and the analog-to-digital converter has $8$-bit precision for liquid crystal driving voltage, so that the grey level of SLMs is from $0$ to $255$. The pixel size of SLMs is $36\,\mu$m$\times36\,\mu$m. The final diffraction pattern is captured on a CMOS camera (CS165MU1 from Thorlabs, Inc.). 

\subsection{Optoelectronic characterizations.} The amplitude modulation response as a function of grey level, which is also corresponding to the applied voltage across a liquid crystal cell, is characterized using a setup shown in \emph{Supplementary Figure\,1b}. The polarizer and analyzer before and after SLMs are configured to be the same as the polarization states used in the CRDONNs experiment setup. The transmitted light intensity is measured by a photodiode (PM16-130 from Thorlabs, Inc.) and the corresponding amplitude is obtained by taking the square root of light intensity. The phase response as a function of grey level is characterized using an interference experiment with a setup shown in \emph{Supplementary Figure\,1c}. The laser beam passes through two identical narrow slits. The polarizer and analyzer before and after the phase-mode SLM (SLMs\,1,\,2,\,3) are configured to have the same polarization states used in the experiment setup. A reference phase mask with the grey level as $0$ is created on one half of the phase-mode SLM, and a measurement phase mask is created on the other half with the grey level changing from $0$ to $255$. An interference pattern is formed on the camera. When the grey level on the measurement phase mask changes from $0$ to $255$, the interference fringes shift. By measuring the fringe shift, the phase response of phase-mode SLM is obtained. 

We use grey levels $0$ and $255$ to represent binarized input images; see detailed description of input image encoding in a following section of \emph{Methods}. Because of the strong amplitude modulation in intensity-mode SLM0 and the coupled modulation of amplitude and phase, we cannot directly measure the phase response using the setup in \emph{Supplementary Figure\,1c}. Instead, we estimate the phase response of SLM0 by matching experimental diffraction patterns with simulation using the setup shown in \emph{Supplementary Figure\,1d}. The detailed description of diffraction models used in simulations is in a following section of \emph{Methods}. The grey level of a small $6\times6$ square region on SLM0 is set as $255$ while the rest is set as grey level $0$. The distance between the intensity-mode SLM and camera is set as $111.76\,$cm. The phase difference $\phi$ between $0$ grey level and $255$ grey level is the only fitting parameter. As shown in \emph{Supplementary Figure\,1d}, the experimentally measured diffraction pattern has a bright spot in the center. Black color in this figure corresponds to large light intensity, while white color corresponds to small light intensity. The red square indicates the captured region on the camera and the square with blue dashed edges indicates an area corresponding to $100\times100$ pixels on the first image SLM. In the simulation, for $\phi=0$, the center spot is dark and as $\phi$ increases to $\sim0.6\pi$ the center becomes bright. The similar patterns can also be observed in the range of $\phi\in[0.6\pi,\pi]$. For simplicity, we associate $\pi$ phase response with the grey level $255$ and we also discover that the diffraction patterns observed in diffractive neural networks are not sensitive to the choice of $\phi$. 

\subsection{Differentiable discrete complex mapping via Gumbel-Softmax.} 
To enable differentiable discrete mapping, our framework defines the input discrete values of grey levels ($V$), which are non-negative integers from $0$ and $255$, as trainable parameters. As shown in Fig.\,\ref{fig:exp_verify}c, the grey level at each pixel of one SLM is represented with a one-hot vector. The experimentally measured electric field amplitude transmission coefficient and phase response as a function of grey level (Fig.\,\ref{fig:exp_verify}b) is a vector of the same length as that of the input grey level vector. In the inference, the amplitude ($t(V)$) and phase ($\phi(V)$) responses are obtained by multiplying the one-hot-format input grey level vector with the measured modulation curves, and the complex-valued number $t(V)\mathrm{e}^{j\phi(V)}$ is used for calculating free-space diffraction; see a following section for a detailed description of diffraction models. Note that the measured modulation curves are different for different SLMs. 

However, one-hot input grey level vectors consisting of discrete values lead to the breakdown of the gradient chain during the backpropagation in training process. To enable backpropagation, we employ the categorical reparameterization via Gumbel-Softmax. Specifically, as shown in Fig.\,\ref{fig:exp_verify}c, during the backpropagation in training process starting from the loss function of ML tasks, instead of propagating to the discrete one-hot voltage levels directly, gradients propagate through the differentiable approximation to the discrete levels generated by Gumbel-Softmax distribution with its class probability $\theta$. The differentiable approximation is updated according to the training algorithm, and the discrete voltage levels are updated by its class probability $\theta$ from approximation. The following describes detailed mathematical formalism. 

Let $T$ be the measured amplitude transmission vector, $\Phi$ be the measured phase response vector, and $V^{m, n}$ be the grey level applied to the pixel located at $(m, n)$ in diffractive layers with a size $N\times N$. The complex-valued transmission $t_{\mathbb{C}}^{m,n}$ provided by the pixel located at $(m, n)$ is expressed in Eq.\,\ref{eq:gumbel-forward}, 
\begin{equation} \label{eq:gumbel-forward}
\begin{split}
\small
V^{m,n} &= \texttt{one\_hot}( \frac{\mathrm{e}^{(\mathrm{log}(\theta^{m,n}) + g^{m,n})/\tau}}{\sum_{1}^{k} \mathrm{e}^{(log(\theta^{i,j}) + g^{i,j})/\tau}}), g^{i,j} \sim \text{Gumbel}(0,1), k = \text{the number of discrete levels}, \\
t_{\mathbb{C}}^{m,n} &=  \overbrace{\underbrace{(V^{m,n} \cdot T)}_{\text{Matmul}} \mathrm{cos}(V^{m,n} \cdot \Phi)}^{\mathrm{Real}} + \underbrace{i(V^{m,n} \cdot A)\mathrm{sin}(V^{m,n} \cdot P)}_{\mathrm{Imaginary}}, \mathrm{for}~m,n \in [0, N-1].
\end{split}
\end{equation}

Let $W$ be a $N\times N$ matrix consisting of elements $V^{m,n}$ with $m,n \in [0, N-1]$. The matrix $W$ has the distribution depending on $\theta$ and the system forward function is $f(W)$. The objective is to minimize the ML task loss $L(\theta) = {\mathbb{E}}_{W \sim p_{\theta}(W)}[f(W)]$, which is the mean squared error (MSE) loss of the image classification in CRDONNs. In the Gumbel-Softmax technique, a Gumbel distribution $g\sim\mathrm{Gumbel}(0, 1)$ and a class probability $\theta$ are introduced to approximate the input discrete grey levels. As a result, any experimentally measured device response consisting of discrete values are differentiable, and can be trained with conventional \texttt{autograd} optimization algorithms to minimize the ML training loss function in CRDONNs. The training process requires us to estimate $\frac{\partial}{\partial{\theta}}\mathbb{E}_{W \sim p_{\theta}(W)} [f(W)]$, which is computed as follows,
\begin{equation} \label{eq:optimization}
\begin{split}
\frac{\partial}{\partial{\theta}}\mathbb{E}_{W\sim p_{\theta}(W)}[f(W)] = \frac{\partial}{\partial {\theta}}\mathbb{E}_{g}[f(G(\theta, g))]  = \mathbb{E}_{g \sim \text{Gumbel}(0, 1)}[\frac{\partial f}{\partial G} \frac{\partial G}{\partial \theta}].
\end{split}
\end{equation}

Specifically, we take the optoelectronic characterization curves of SLMs shown in Fig.\,\ref{fig:exp_verify}b in the training process of our CRDONNs system, where the trainable parameters, input discrete grey levels for each pixel in SLM, are represented with a $1\times256$ one-hot vector, and experimentally measured amplitude or phase response is in the shape of $256\times1$. Each SLM layer in our system has a size of $260\times260$ (size $260$ includes $80$ paddings on each side of SLMs), and thus the total number of trainable parameters in each layer is $260\times260\times256$. The distance between layers is $27.94\,$cm. For demonstrated three-labels models, three detector regions with a size of $10\times10$ are pre-defined, and the sums of the intensity of these regions are represented in a $1\times10$ vector in \texttt{float32} type. The final prediction results are generated after going through the operation of \texttt{argmax}. The MSE loss is used to guide the training process, which is iteratively optimized via the automatic optimizer \texttt{Adam}. The number of training epochs is $100$, the learning rate is $0.5$, and the batch size is $500$. All the training calculations were done on NVIDIA 3090 Ti GPUs. 

\subsection{Input image encoding.} The original input images from \texttt{MNIST}\cite{LeCunEtAl1998HLC} and \texttt{F-MNIST}\cite{XiaoEtAl2017APA} with size of $28 \times 28$ are first interpolated into a size of $100\times100$ and then binarized. Specifically, the pixels with the normalized brightness greater than $0.5$ are encoded corresponding to the grey level $255$ and the rest pixels are encoded corresponding to the grey level $0$. The intensity-mode image SLM has an light intensity extinction ratio $\sim1000$ for two extreme grey levels, corresponding to the electric field amplitude extinction ratio $\sim31.6$. In addition, there is a $\pi$ phase difference between $0$ and $255$ grey levels. Thus, the complex-valued field transmission becomes $0.0316$ for $0$ grey level and dark pixels in input images, and $e^{i\pi} = -1$ for $255$ grey level and bright pixels in input images. 

\subsection{Free-space diffraction calculations.} The scalar diffraction theory is mainly used in describing free-space diffraction between SLMs and the camera; see \emph{Supplementary Figure\,2} for an illustration. Specifically, the input at point $(x,y)$ on $l$-th layer can be written as the summation of all the outputs at $(l-1)$-th layer as
\begin{equation} \label{eq:free-space-prop}
f_l(x,y,z)={\iint}f_{l-1}(x',y',0)h(x-x',y-y')dx'dy',
\end{equation}
where $z$ is the distance between two diffractive layers, $h$ is the impulse response function of free space, $f_{l-1}$ is the output wavefunction of points on $(l-1)$-th layer and also the input wavefunction of free space propagation, $f_l$ is the output function of free-space propagation and also the input for the phase mask at $l$-th plane. There are different ways of calculating Eq.\,\ref{eq:free-space-prop}. 

The first method of calculating the integral in Eq.\,\ref{eq:free-space-prop} is based on a spectral algorithm. Specifically, the integral is expressed as the convolution of $f_{l-1}$ and $h$ and with the assist of Fast Fourier Transform (FFT) under \texttt{Pytorch} framework, this convolution can be calculated in a fast and trainable manner\cite{pytorch_fft}. By convolution theorem, the two-dimensional (2D) Fourier transformation over $x,y$-axes of the convolution of $f_{l-1}$ and $h$ is the product of 2D Fourier transformations over $x,y$-axes of $f_{1-1}$ and $h$ through
\begin{equation}
\mathcal{F}_{xy}(f_l(x,y,z))=\mathcal{F}_{xy}(f_{l-1}(x,y,0))\mathcal{F}_{xy}(h(x,y)), 
\end{equation}
\begin{equation}\label{eq_fourier}
    U_l(\alpha,\beta, z)=U_{l-1}(\alpha,\beta, 0)H(\alpha,\beta),
\end{equation}
where $U$ and $H$ are the Fourier transformation of $f$ and $h$, respectively. To calculate the free-space propagation between diffractive layers, the input signal and the free-space impulse function is first converted to spatial frequency domain ($\alpha, \beta$) using FFT. The output of free-space propagation is simply the product of the input FFT signal and free space transfer function $H$, which is then converted back to spatial domain ($x,y$) through inverse FFT. The impulse function we use is the Fresnel approximation with 
\begin{equation}
  h(x,y)=\frac{\mathrm{exp}(ikz)}{i\lambda z}\mathrm{exp}\{\frac{ik}{2z} (x^2+y^2)\}, 
\end{equation}
where $k = 2\pi/\lambda$ is free-space wavenumber with the wavelength $\lambda=532\,$nm. 


The second method of calculating the integral in Eq.\,\ref{eq:free-space-prop} is the direct integral as a convolution. Following the Fresnel-Kirchhoff diffraction formula, the Eq.\,\ref{eq:free-space-prop} can be re-written as
\begin{equation} \label{eq:free-space-prop-2}
  f_l(x,y,z)={\frac{k}{2\pi i z}}{\iint}f_{l-1}(x',y',0)\mathrm{exp}\{ik\frac{(x-x')^2 + (y-y')^2}{2z}\}dx'dy'.
\end{equation} 
The detailed calculation process can be found in Ref.\,\cite{VdovinEtAl1997}. A simplified one-dimensional version is described here. Let the integral be defined in a finite interval $[-L/2, L/2]$,
\begin{equation} \label{eq:free-space-prop-3}
  f_l(x,z)=\sqrt{\frac{k}{2\pi i z}}{\int_{-L/2}^{L/2}}f_{l-1}(x',0)\mathrm{exp}\{ik\frac{(x-x')^2}{2z}\}dx'.
\end{equation}
The functions $f_l(x)$ and $f_{l-1}(x')$ are replaced with step functions $U_s$ and $U_t$ defined in a grid with $s \in[1,N]$ and $t\in[1,N]$. Thus, the Eq.\,\ref{eq:free-space-prop-3} is converted to
\begin{gather} 
  U_j=\sqrt{\frac{k}{2\pi i z}}\Sigma_{2}^{N-1}U_s\int_{x_s-0.5}^{x_s+0.5}\mathrm{exp}\{ik\frac{(x_j - x)^2}{2z}\} + \nonumber \\U_1\int_{x_1}^{x_1+0.5}\mathrm{exp}\{ik\frac{(x_j - x)^2}{2z}\} + U_N\int_{x_N-0.5}^{x_N}\mathrm{exp}\{ik\frac{(x_j - x)^2}{2z}\}.
  \label{eq:free-space-prop-4}
\end{gather}
We analytically calculate the integral in Eq.\,\ref{eq:free-space-prop-4} using the Fresnel integral functions provided in the \texttt{scipy} package and Eq.\,\ref{eq:free-space-prop-4} becomes
\begin{equation} 
  \label{eq:free-space-prop-5}
  U_j= \Sigma_{2}^{N-1}U_sK_{js} + U_1K_{j1} + U_NK_{jN},
\end{equation}
which can also be evaluated using FFT for fast calculations. We refer this calculation method as convolutional Fresnel method. The comparison of simulation results trained and inferenced using different methods together with experimental results are summarized in \emph{Supplementary Figure\,3}.


\subsection{Physics-aware complex-valued FGSM (C-FGSM).} To explore the behavior of the CRDONNs system with Gumbel-Softmax under adversarial attacks, we employ C-FGSM to generate adversarial images, which can be recognized correctly by human eyes but can confuse the prediction model. Specifically, in the physics-aware optical system, we have the input image interpolated and binarized. The pixel value for featured part in the image is $-1$ (grey level $255$), where the phase for the light wave is $\pi$ and the amplitude of the light wave is $1$. The rest is $0.0316$ (grey level $0$), where the phase for the light wave is $0$ and the amplitude of the light wave is $0.0316$. As a result, adversarial attacks are binarized attacks. With C-FGSM, the original examples ($I_p$) before attack is first evaluated with the trained model and the loss function ($\mathcal{L}(I_p,I_t)$) is computed with the ground truth ($I_t$). In the computation graph for the model, we obtain the gradients applied to input examples with respect to the loss function. Note that the CRDONNs system is described with complex-valued numbers in simulations, the obtained gradients from simulations are also complex-valued numbers. In an optimization process, we utilize descent gradients to minimize the loss function while in an adversarial generation, we utilize ascent gradients at the input examples to increase the loss function and confuse the model. Considering the encoding method for the input images in the CRDONNs system, we only apply the real parts of the gradients obtained from simulations to input images. Therefore, in binary examples, we flip the value of an image pixel when the corresponding applied gradient has a positive sign, otherwise it remains the same. Additionally, to make sure the adversarial examples are readable for human eyes, we apply a mask on the featured part in input images, which is the part with pixel value of $-1$ in original images, to disable the application of attack gradients to image features.

\subsection{MLPs and CNNs architectures.} 
In the comparative analysis of adversarial attacks on different neural networks, the MLPs model consists of two linear layers with hidden size $128$, where the input image is flattened as one-dimensional tensor, i.e., MLP ($67600 \rightarrow 128 \rightarrow 10$). The CNNs model consists of two Conv2D, two MaxPooling2D, two linear layers that make the final predictions. Specifically, first Conv2D is configured with kernel size $(5,5)$, $32$ filters, with stride and padding being $2$; second Conv2D layer has the same configuration as first Conv2D, except that the number of filters increases to $64$; both MaxPooling layers have $(3,3)$ kernels with the stride set as $2$. The two linear layers are set as $64\times15\times15\,(14400) \rightarrow 256 \rightarrow 10$, in order to match the output tensor size of convolution layers.

\end{methods}

\pagebreak
\begin{addendum}


\item [Data availability] All other data that support the plots within this paper and other findings of this study are available from the corresponding authors upon reasonable request. 

\item [Code availability] All codes are available from the corresponding authors upon reasonable request. 
    
\item [Acknowledgements] R.\,C., J.\,F., and W.\,G. thank the support from the University of Utah start-up fund. Y.\,L. and C.\,Y. thank the support from grants NSF-2019336 and NSF-2008144. 

\item [Author contributions] B.\,S.\,R., C.\,Y., and W.\,G. conceived the idea and designed the project. R.\,C. built the experimental setup and performed experiments under the supervision and guidance of W.\,G. Y.\,L. built the simulation software and performed machine learning calculations and analysis under the supervision and guidance of C.\,Y. M.\,L., J.\,F., and Y.\,T. help with both experimental setup and simulation programs. C.\,Y. and W.\,G. wrote the manuscript. All authors discussed the manuscripts and provided the feedback. 

\item[Competing Interests] The authors declare that they have no competing financial interests.

\item[Correspondence] Correspondence and requests for materials should be addressed to Cunxi Yu (email: cunxi.yu@utah.edu) and Weilu Gao (email: weilu.gao@utah.edu).
\end{addendum}

\newpage

\begin{table}
  \centering
  \caption{\texttt{MNIST} accuracy before and after adversarial attacks}
  \label{table:mnist_acc}
  \begin{tabularx}{\textwidth}{|| X || X | X ||}
    \hline
    3-label model & Accuracy before attack & Accuracy after attack \\ [0.5ex] 
    \hline\hline
    $\{0,1,2\}$ & {98\,\%} & {63\,\%}\\
    \hline
    $\{0,3,7\}$ & {98\,\%} & {81\,\%}\\
    \hline
    $\{0,5,9\}$ & {97\,\%} & {77\,\%}\\
    \hline
    $\{1,5,8\}$ & {95\,\%} & {41\,\%}\\
    \hline
    $\{3,4,5\}$ & {93\,\%} & {49\,\%}\\
    \hline
    \hline
  \end{tabularx}
  \\
\end{table}

\newpage
\bibliography{weilu.bib} 

\end{document}